\documentclass[12pt]{article}

\usepackage{amsmath,amssymb,epsfig}
\usepackage{graphicx}
\usepackage{booktabs}

\begin{document}
\setlength{\parskip}{0.45cm}
\setlength{\baselineskip}{0.75cm}

%
%
%
\begin{titlepage}
\setlength{\parskip}{0.25cm}
\setlength{\baselineskip}{0.25cm}
\begin{flushright}
DO-TH 06/04\\
\vspace{0.2cm}
June 2006
\end{flushright}
\vspace{1.0cm}
\begin{center}
\large
{\bf Parton evolution in the fixed flavor factorization scheme}
\vspace{1.5cm}

\large
M.~Gl\"uck and E.~Reya\\
\vspace{1.0cm}

\normalsize
{\it Institut f\"{u}r Physik´, Universit\"{a}t Dortmund}\\
{\it D-44221 Dortmund, Germany} \\
\vspace{0.5cm}

\vspace{1.5cm}
\end{center}

\begin{abstract}
\noindent It is argued that while the scale dependence of the parton
distributions in the fixed flavor factorization scheme is governed
by three active flavors, the scale dependence of the running
coupling should nevertheless be better governed by a {\em{variable}} number 
of active flavors.
\end{abstract}
\end{titlepage}


The fixed flavor factorization schme (FFS) is characterized by considering
the heavy quarks ($h=c,\, b,\, t$) always as external particles which are
not included among the partons in the colorless hadrons. Their 
participation in deep inelastic scattering processes like, say,
$eN\to eX$ is considered to be due to production subprocesses such 
as $\gamma^*g\to h\bar{h}$ rather than $\gamma^*h\to h$.  The latter
subprocess becomes relevant in the so called variable flavor
factorization scheme where, besides the light $u$, $d$, $s$ quarks,
the heavy quarks are also considered to form an intrinsic part of
colorless hadrons. Considering both subprocesses together would
amount to double counting and thus the FFS dictates setting the
number of flavors $n_f=3$ in the flavor--singlet QCD evolution
equations
\begin{equation}
\frac{d}{d\ln Q^2}\, \vec{q}\, (x,Q^2) = \frac{\alpha_s(Q^2)}{2\pi}\,
  \int_x^1\, \frac{dy}{y}\, \hat{P}\left( \frac{x}{y},Q^2\right)\, 
    \vec{q}\, (y,Q^2)
\end{equation}
where $\vec{q}=(\Sigma,g)^T$ with
\begin{equation}
\Sigma(x,Q^2) = \sum_{q=u,d,s}(q+\bar{q})
\end{equation}
and $\hat{P}(z,Q^2) = \hat{P}^{(0)}(z)+\frac{\alpha_s}{2\pi}\hat{P}^{(1)}
(z)+ (\frac{\alpha_s}{2\pi})^2 \hat{P}^{(2)}(z)$.  In order to keep our arguments
transparent as far as possible, we neglect, without loss of generality,
the NLO (2--loop) and NNLO (3--loop) splitting functions $\hat{P}^{(1)}$
and $\hat{P}^{(2)}$, respectively, and the well known leading order (LO)
splitting functions are given by \cite{ref1}
\begin{equation}
\hat{P}^{(0)} = \left( \begin{array}{cc}
P_{qq}^{(0)}, & 2n_fP_{qg}^{(0)}\\
P_{gq}^{(0)}, & P_{gg}^{(0)}\end{array} \right)
\end{equation}
with $P_{qq}^{(0)}(z) = C_F\,\left(\frac{1+z^2}{1-z}\right)_+$, 
$P_{qg}^{(0)}(z) = T_R\left[z^2+(1-z)^2\right]$, $P_{gq}^{(0)}(z) =
C_F\left[1+(1-z)^2\right]/z$
and
\begin{equation}
P_{gg}^{(0)}(z) = 2\, C_A\left[\frac{z}{(1-z)}_+ +\frac{1-z}{z} +z(1-z)\right]
 + c_{\delta}\,  \delta(1-z)
\end{equation}
where $C_F=\frac{4}{3}$, $C_A = 3$ and $T_R=\frac{1}{2}$.  The $n_f$ dependence
of $P_{gg}^{(0)}(z)$ resides in the endpoint ($z=1$) contribution
$c_{\delta} =\frac{11}{6}\, C_A-\frac{2n_f}{3}\, T_R$ which derives from
the energy--momentum sum rule
\begin{equation}
\int_0^1 \left[ x\Sigma(x,Q^2) + xg(x,Q^2) \right] \, dx = 1\, \, ,
\end{equation}
dictating the  `second moment' constraints
\begin{eqnarray}
\int_0^1 z \left[P_{qq}^{(0)}(z) + P_{gq}^{(0)}(z)\right] 
                      \, dz & = & 0\nonumber\\
\int_0^1 z \left[ 2n_f P_{qg}^{(0)}(z)+P_{gg}^{(0)}(z)\right]
                      \, dz & = & 0\,\,  .
\end{eqnarray}
Inserting (4) into this latter constraint gives $2n_f T_R \frac{1}{3} +
(-\frac{11}{6} C_A+c_{\delta}) = 0$, i.e.\ $c_{\delta}$ given above.
In other words, the $n_f$ dependence of $P_{gg}^{(0)}$ derives, as is 
well known \cite{ref1}, from the
`second moment' of the splitting function $2n_f P_{qg}^{(0)}$ describing
the splitting of the gluon $g\to q\bar{q}$ into $n_f$ massless on--shell
quark--antiquark pairs. The choice $n_f=3$ in (3) is dictated by the FFS
expression for $\Sigma(x,Q^2)$ in (2).  Consequently, fixing $n_f=3$ in
$P_{gg}^{(0)}$ in (4) is needed to guarantee the $Q^2$ independence of
the energy--momentum sum rule (5).

On the other hand, the running coupling $\alpha_s(Q^2)$ in (1) evolves
according to
\begin{equation}
\frac{d\alpha_s(Q^2)}{d\ln Q^2} = \beta(\alpha_s) = -\beta_0\, 
                       \frac{\alpha_s^2(Q^2)}{4\pi} +\ldots
\end{equation}
where
\begin{equation}
\beta_0 = 11-\frac{2}{3} n_f
\end{equation}
and $n_f$ is the number of active quark flavors $f$ satisfying, in the 
most commonly used $\overline{\rm MS}$ renormalization scheme, 
$m_f^2\leq Q^2$.
As is well known, this $n_f$ dependence in (8) derives from the same
fermionic 1-loop vacuum polarization diagram as in QED, by properly
taking into account the color degrees of freedom.  Furthermore, it is
the scale (virtuality) $Q^2$ which dictates how many quark flavors 
effectively contribute in (7).  The question arises whether
the FFS also dictates \cite{ref2} setting always $n_f=3$ in $\beta_0$.  
First of all one
notes that this is not needed in order to guarantee the $Q^2$ independence
of (5).  Secondly the inclusion of the heavy quark $h$--loops in the 
vacuum polarization diagrams, responsible for the $n_f$ {\em dependence}
in $\beta_0$, does not involve double counting since this contribution
is not accounted for in the FFS as explained above.  For this second
reason one may keep the $n_f$ dependence in $\beta_0$, i.e.\ in 
$\alpha_s(Q^2)$, in contrast to the situation in (3) and (4) where it
would lead to {\em double} counting in the FFS since here the heavy
quark flavors $h = c,\, b,\, t$ are already taken care of via the 
production process $\gamma^*g\to h\bar{h}$
as argued above.

It is, of course, also possible to fix $n_f=3$ in the $\beta$--function
governing the evolution of $\alpha_s(Q^2)$ as was considered in 
\cite{ref2}.  But then one encounters large higher order 
$\alpha_s^{(n_f=3)}(Q^2)\ln\frac{Q^2}{m_h^2}$ logarithmic corrections
in all subsequent calculations like, e.g., the $Q^2$--evolution 
equations for the u,d,s and gluon distributions, which {\em must} be
taken into account.
For this reason one should better choose for the calculation of these
distributions the variable flavor number scheme in the $\beta$--function,
even in the so called fixed flavor factorization scheme, a practice
followed in many analyses carried out within this factorization scheme.
Making this choice, automatically resums the above mentioned higher order
logarithmic corrections and consequently improves the stability of the
perturbative expansion.  At any rate, simply fixing \cite{ref2} $n_f=3$
in $\beta_0$ and $\beta_1$ does not generate the correct $Q^2$--evolution
of the u,d,s and $g$ distributions.
A possible objection to the variable flavor scheme in $\beta(\alpha_s)$
could be that the NLO coefficient functions for the $c\bar{c}$ pair
production have been calculated \cite{ref3} in a renormalization scheme where
the coupling $\alpha_s$ evolves according to $n_f=3$. However, as 
demonstrated in \cite{ref4}, it is perfectly appropriate to choose
$\mu_r=\mu_f={\cal{O}}(m_c)$ for the renormalization and factorization scales in
calculating the NLO $c\bar{c}$ pair production cross section, i.e.,
a scale $\mu_r$ where $n_f=3$ in $\beta(\alpha_s)$ is the appropriate
choice.

Let us finally note that the argument in \cite{ref2} for fixing
$n_f=3$ in $\beta_0$ is actually not conclusive.  This can be easily
seen by considering the LO quark contribution to the longitudinal
structure function $F_L$ which results in an $n_f$ dependence of
$\partial F_L/\partial \ln Q^2$ in contrast to the situation presented
in \cite{ref2} where only the gluon contribution to $F_L$ was taken into
account.  Furthermore, it is well known that also $\partial F_2/\partial
\ln Q^2$ is $n_f$ dependent.  Thus the demand \cite{ref2} for the 
$n_f$ independence of these quantities does not provide any meaningful
criterion and can moreover never be realized.

To summarize, in contrast to the assertions in \cite{ref2}, we
conclude that it is neither wrong nor inconsistent to choose a scale
dependent $n_f$ in $\beta(\alpha_s)$ within the fixed flavor 
factorization scheme.
The arguments presented here, mainly within the LO framework, 
hold obviously also for the higher perturbative
orders, but their explicit demonstration is technically more involved
and consequently less transparent.
\vspace{2.0cm}




\end{document}